\documentclass[%
 reprint,
 amsmath,amssymb,
 aps,
]{revtex4-1}

\usepackage{graphicx}
\usepackage{dcolumn}
\usepackage{bm}

\newcommand{\be}{\begin{equation}}
\newcommand{\ee}{\end{equation}}
\newcommand{\ba}{\begin{eqnarray}}
\newcommand{\ea}{\end{eqnarray}}
\newcommand{\baa}{\begin{eqnarray*}}
\newcommand{\eaa}{\end{eqnarray*}}
\newcommand{\bb}{\begin{thebibliography}}
\newcommand{\eb}{\end{thebibliography}}

\newcommand{\bi}[1]{\bibitem{#1}}
\newcommand{\lab}[1]{\label{#1}}
\newcommand{\re}[1]{(\ref{#1})}

\renewcommand\t{\tilde}

\begin{document}

\preprint{APS/123-QED}

\title{How to construct spin chains with perfect state transfer}

\author{Luc Vinet}
 \affiliation{Centre de recherches math\'ematiques
Universit\'e de Montr\'eal, P.O. Box 6128, Centre-ville Station,
Montr\'eal (Qu\'ebec), H3C 3J7}
\author{Alexei Zhedanov}%
\affiliation{Institute for Physics and Technology, R.Luxemburg
str. 72,  83114, Donetsk, Ukraine }


\begin{abstract}
It is shown how to systematically construct the $XX$ quantum spin
chains with nearest-neighbor interactions that allow perfect state
transfer (PST). Sets of orthogonal polynomials (OPs) are in
correspondence with such systems. The key observation is that for
any admissible one-excitation energy spectrum, the weight function
of the associated OPs is uniquely prescribed. This entails the
complete characterization of these PST models with the mirror
symmetry property arising as a corollary. A simple and efficient
algorithm to obtain the corresponding Hamiltonians is presented. A
new model connected to a special case of the symmetric $q$-Racah
polynomials is offered. It is also explained how additional models
with PST can be derived from a parent system by removing energy
levels from the one-excitation spectrum of the latter. This is
achieved through Christoffel transformations and is also
completely constructive in regards to the Hamiltonians.

\end{abstract}



\maketitle

\section{Introduction}
The problem of perfect state transfer
(PST) in quantum information processing is deservedly attracting
much attention. (See Refs. \cite{Bose}, \cite{Kay} for reviews.)
The transport of quantum state from one location to another is
perfect if it is realized with probability 1, thereby avoiding
dissipation. Occurrences of perfect transmission have been found
in some $XX$ chains with inhomogeneous couplings \cite{Bose},
\cite{Kay}, \cite{Albanese}, \cite{Karbach}: in these cases, the
probability for the transfer of a single spin excitation from one
end of the chain to the other is indeed found to be 1 for certain
times. These models have the advantage that the perfect transfer
can be done without the need for active control. We shall here
describe how such systems can be systematically "constructed" from
given one-excitation spectra.

Specifically, we shall focus on Hamiltonians $H$ of the $XX$ type
with nearest-neighbor interactions, i.e. \be H=\frac{1}{2} \:
\sum_{l=0}^{N-1} J_{l+1}(\sigma_l^x \sigma_{l+1}^x + \sigma_l^y
\sigma_{l+1}^y) +  \frac{1}{2} \: \sum_{l=0}^N B_l(\sigma_l^z +1),
\lab{H_def} \ee where $J_l$ are the constants coupling the sites
$l-1$ and $l$ and $B_l$ are the strengths of  the magnetic field
at the sites $l$ ($l=0,1,\dots,N$). The symbols $\sigma_l^x, \:
\sigma_l^y,\: \sigma_l^z$ stand for the Pauli matrices which act
as follows on the single qubit states $| 0 \rangle$ and $| 1
\rangle$:
$$
\sigma^x | 0 \rangle = | 1 \rangle, \quad \sigma^y | 0 \rangle = -i | 1 \rangle, \quad \sigma^z | 0 \rangle = -| 0 \rangle
$$
$$
\sigma^x | 1 \rangle = | 0 \rangle, \quad \sigma^y | 1 \rangle = i | 0 \rangle, \quad \sigma^z | 1 \rangle = | 1 \rangle,
$$
It is straightforward to see that
$$
[H, \frac{1}{2} \: \sum_{l=0}^N (\sigma_l^z +1)]=0
$$
and therefore that the eigenstates of $H$ split in subspaces
labeled by  the number of spins over the chain that are in state
$| 1 \rangle$. For our purposes, it will suffice to restrict $H$
to the subspace spanned by the states witch contain only one
excitation (or spin up). A natural basis for that subspace is
given by the vectors
$$
|e_n ) = (0,0,\dots, 1, \dots, 0), \quad n=0,1,2,\dots,N,
$$
where the only "1" occupies the $n$-th position.  In that basis,
the restriction $J$ of $H$ to the one-excitation subspace is given
by the following $(N+1) \times (N+1)$  Jacobi matrix
\[
J =
 \begin{pmatrix}
  B_{0} & J_1 & 0 &    \\
  J_{1} & B_{1} & J_2 & 0  \\
   0  &  J_2 & B_2 & J_3     \\
   &   &  \ddots &    \ddots  \\
    & & \dots & J_N & B_N
\end{pmatrix},
\]
where it is assumed that $J_i>0$. Its action on the basis vectors
$|e_n)$ reads \be J |e_n) = J_{n+1} |e_{n+1}) + B_n |e_n) + J_{n}
|e_{n-1}). \lab{Je} \ee Note also that the conditions \be
J_0=J_{N+1}=0 \lab{J0} \ee are assumed.

It has been shown \cite{Kay} that the eigenvalues of $J$ must
satisfy a simple  PST condition for the perfect state transfer to
be possible. (We shall discuss this in Section 3.) Since they
encode 3-term recurrence relations, the Jacobi matrices are
diagonalized by orthogonal polynomials (OPs). Spin chains allowing
PST are hence in correspondence with families of OPs.

One question we address here is the following. Given spectral data
sets satisfying the  PST condition, can the corresponding spin
chains with the PST property be found? In other words, is there a
procedure to explicitly obtain the parameters $J_n$ and $B_n$ that
determine the Hamiltonians? As it turns out this question has a
rather elegant affirmative answer. Indeed, we shall see that the
condition for a spin chain to possess the PST property is very
simply expressed through the weight function of the orthogonal
polynomials associated to the chain. Hence, given a set of
eigenvalues satisfying the PST condition, the weight function is
then prescribed. This uniquely fixes (up to normalization) the
corresponding orthogonal polynomials and therefore the
coefficients $J_n$ and $B_n$ as their recurrence coefficients. A
simple algorithm for constructing spin chains with the PST
property along these lines will be presented.

Another constructive issue that we consider is this: can one
obtain  different spin chains with the PST property by performing
some appropriate surgery on the spectrum of a model already known
to generate PST? Here again the answer is  positive and also
provides tools to explicitly construct many spin chains with PST
from a given one.

These two natural questions have been touched upon in
\cite{Karbach} and \cite{Wang}.  However, the link with the weight
function of the associated OPs had so far not been stressed and
therein lie the constructive power that our considerations add.

The outline of the paper is as follows.

In Section 2 we briefly describe  standard results concerning Jacobi matrices and orthogonal polynomials.

In Section 3, we revisit the necessary and sufficient conditions
for  $XX$ spin chains to realize perfect state transfer. We derive
the spectral condition already referred to and obtain the
expression for the weight function of the associated polynomials
on which much of the model characterization hinges.

In Section 4, we offer a simple algorithm for constructing the
Hamiltonians of the chains with PST from the spectra and the
weight functions.

In Section 5, we give two simple examples that illustrate how the
proposed method applies.  The first is the well known case
corresponding to a linear spectrum and the Krawtchouk polynomials.
The second stems from a hyperbolic energy spectrum and leads to a
novel spin chain Hamiltonian with PST that can be analytically
described.

In Section 6, we present a "surgical" procedure consisting in the
removal of energy levels from the one-excitation spectrum of a
spin chain with PST.  It is shown to lead to other spin chains
with the same property. This spectral surgery amounts to
performing Christoffel transforms of the orthogonal polynomials
associated to the parent or original system. Since the transformed
polynomials are explicitly known (from the general OP theory
\cite{Chi}), such is also the case for the corresponding Jacobi
matrix and spin chain Hamiltonian.

\section{Finite Jacobi matrices and orthogonal polynomials}
Because the matrix $J$ is Hermitian, there exists an eigenbasis $| s \rangle$ such that
\be
J | s \rangle = x_s | s \rangle, \quad s=0,1,\dots,N . \lab{J_eig} \ee
The eigenvalues $x_s$ are all real and nondegenerate $x_s \ne x_t$ if $s \ne t$.

Consider the expansion of the eigenbasis $| s \rangle$ in terms of
the basis $|e_n)$ \be | s \rangle = \sum_{n=0}^N W_{sn} |e_n) .
\lab{se_expans} \ee From \re{J_eig} it is seen that the expansion
coefficients $W_{sn}$ satisfy \be J_{n+1} W_{s,n+1} + B_n W_{sn} +
J_{n} W_{s,n-1} = x_s W_{sn} . \lab{rec_W} \ee Hence we can
present them in the form \be W_{sn} = W_{s0} \chi_n(x_s),
\lab{W_pi} \ee where $\chi_n(x)$ are polynomials satisfying the
recurrence relation \be J_{n+1} \chi_{n+1}(x) + B_n \chi_n(x) +
J_{n} \chi_{n-1}(x) = x \chi_n(x) \lab{rec_pi} \ee and the initial
conditions \be \chi_0=1, \; \chi_{-1}=0 . \lab{ini_pi} \ee Both
bases $|e_n)$ and $|s\rangle$ are orthonormal
$$
(e_m|e_m) =\delta_{nm}, \quad  \langle s' | s\rangle =\delta_{ss'}.
$$
Whence the matrix $W_{sn}$ is orthogonal
\be
\sum_{n=0}^N W_{sn}W_{s' n} = \delta_{s s'} \lab{ort_W} \ee
and similarly
\be
\sum_{s=0}^N W_{sn}W_{s m} = \delta_{nm}. \lab{ort_W_1} \ee
From \re{ort_W_1} it follows that the polynomials $\chi_n(x)$ are orthonormal on the finite set of spectral points $x_s$
\be
\sum_{s=0}^N w_s \chi_n(x_s) \chi_m(x_s) = \delta_{nm}, \lab{ort_pi} \ee
where
\be
w_s = W_{s0}^2 \lab{wW} \ee
play the role of discrete weights for the polynomials $\chi_n(x)$.

Note that
\be
\sum_{s=0} w_s = \sum_{s=0} W_{s0}^2 =1 \lab{norm_w} \ee
which amounts to the standard normalization condition for the weights.

We thus have the expansions
\be
| s \rangle = \sum_{n=0}^N \sqrt{w_s} \chi_n(x_s) |e_n) \lab{sew_expans} \ee
and similarly
\be
| e_n ) = \sum_{s=0}^N \sqrt{w_s} \chi_n(x_s) |s \rangle . \lab{esw_expans} \ee
In what follows it will be convenient to work with  monic orthogonal polynomials
\be
P_n(x) = J_1 J_2 \dots J_n \: \chi_n(x) = x^n + O(x^{n-1}). \lab{P_n_def} \ee
They satisfy the recurrence relation
\be
P_{n+1}(x) + B_n P_n(x) + U_n P_{n-1}(x) = xP_n(x), \lab{recP} \ee
where $U_n = J_n^2>0$.

The polynomials $P_n(x)$ satisfy the orthogonality relation
\be
\sum_{s=0}^N P_n(x_s) P_m(x_s) w_s = h_n \: \delta_{nm}, \lab{ort_P} \ee
where
$$
h_n = U_1 U_2 \dots U_n .
$$
Starting from the initial conditions $P_0=1, P_{-1}=0$, it is
possible to reconstruct all monic polynomials $P_n(x),
n=1,2,\dots,N$ uniquely.  The polynomial $P_{N+1}(x)$ will be \be
P_{N+1}(x) = (x-x_0)(x-x_1) \dots (x-x_N).  \lab{P_N+1} \ee From
the standard theory of orthogonal polynomials \cite{Chi} it
follows that the discrete weights can be expressed as \be w_s =
\frac{h_N}{P_N(x_s) P_{N+1}'(x_s)}, \quad s=0,1,\dots, N
.\lab{w_s_PP} \ee In what follows we shall take the eigenvalues
$x_s$ to be in increasing order i.e. we shall assume that \be
x_0<x_1<x_2 < \dots <x_N . \lab{incr_x} \ee Such an ordering is
always possible because, by assumption, all eigenvalues $x_s$ of
the Hermitian matrix $J$ are real and simple.

One can then present the expression for $P_{N+1}'(x_s)$ in the
form \ba &&P_{N+1}'(x_s) = (x_s-x_0)(x_s-x_1) \dots \times
\lab{N+1_prod} \\&&(x_s-x_{s-1}) (x_s-x_{s+1}) \dots (x_s-x_N) =
\nonumber \\&& (-1)^{N+s} |P_{N+1}'(x_s)|.
 \nonumber \ea

\section{Necessary and sufficient conditions for perfect quantum state transfer}
A perfect quantum state transfer is
realized if for some fixed time $T>0$ one has \be e^{iTJ} |e_0) =
e^{i \phi} |e_N ), \lab{pqc} \ee where $\phi$ is a real number. In
other words, the initial state $|e_0 )$ evolves over time $T$ into
the state $|e_N )$ (up to an inessential phase factor $e^{i
\phi}$).

Expanding the states $|e_0)$ and $|e_N )$ in terms of the
eigenstates $| s \rangle$ with the help of formula \re{esw_expans}
we find that condition \re{pqc}  is equivalent to \be \chi_N(x_s)
= e^{-i \phi} e^{i T x_s}, \quad s=0,1,\dots, N . \lab{pqc_1} \ee
Now, on the one hand, since the polynomial $\chi_N(x)$  is real,
it follows from \re{pqc_1} that \be \chi_N(x_s) = \pm 1.
\lab{pqc_2} \ee On the other hand, the orthogonal polynomials
$\chi_n(x)$ possess the interlacing property \cite{Chi}. In
particular, any zero
 of the polynomial $\chi_N(x)$ should lie between two neighboring zeroes $x_s$ of the polynomial $\chi_{N+1}(x)$.
 This is possible only if the values $1$ and $-1$ alternate in \re{pqc_2},
 namely if
 \be
 \chi_N(x_s) = (-1)^{N+s}, s=0,1,\dots, N \lab{pqc_3} \ee
where we took into account the ordering \re{incr_x}, formula \re{N+1_prod} and the positivity of the weights $w_s$ in formula \re{w_s_PP}.

From \re{pqc_1} it follows that \re{pqc_3} is equivalent to the
condition \be e^{-i \phi} e^{i T x_s} = e^{i \pi s} e^{i \phi +
\pi (N+2G_s)} \lab{pqc_4} \ee where $G_s$ are arbitrary integers.

Whence we have the following condition for the spacings between
the successive levels \be x_{s+1} - x_s= \frac{\pi}{T} M_s,
\lab{xxM} \ee where $M_s$ may be arbitrary positive odd numbers.

We thus derived the necessary conditions \re{xxM} and \re{pqc_3}
for a spin chain to realize perfect state transfer.

It is easily seen that these conditions are also sufficient.

From \re{pqc_3}, \re{w_s_PP} and \re{N+1_prod} it is observed that
\re{pqc_3} is equivalent to the condition \be w_s =
\frac{\kappa_N}{|P_{N+1}'(x_s)|}>0 \lab{w_s_kap} \ee where the
normalization constant $\kappa_N$ (chosen from the condition $w_0
+ w_1 + \dots+w_N=1$) is not essential for our considerations.
This formula can also be found in \cite{BG} where it occurs in
connection with the inverse spectral problem for persymmetric
matrices. (See below.)

This means that only condition \re{xxM} is crucial: once the
eigenvalues $x_s$ satisfy \re{xxM}, the weights $w_s>0$ are then
uniquely determined (up to a common factor) via \re{w_s_kap} and
\re{N+1_prod}.  In turn, the spectral points $x_s$ together with
the weights $w_s$ are known to determine the Hermitian Jacobi
matrix $J$ uniquely \cite{Chi}.  Hence, from the sole knowledge of
the spectral points $x_s$ (satisfying condition \re{xxM}) we can
uniquely reconstruct the spin chain Hamiltonians with the desired
properties.

\section{Reconstruction of the matrix $J$ from spectral data}
Consider now the matrix $J^*$ which is
obtained from $J$ by a reflection with respect to the main
counter-diagonal, i.e. \be J^* =RJR, \lab{J*J} \ee where the
matrix $R$ (reflection matrix) is
\[
R=\begin{pmatrix}
  0 & 0 & \dots & 0 & 1    \\
  0 & 0 & \dots  & 1 & 0  \\
    \dots  & \dots & \dots & \dots & \dots      \\
   1 & 0 &  \dots & 0 &0  \\

\end{pmatrix}.
\]
The matrix $J^*$ has the same structure as the matrix $J$ (i.e. $J^*$ is a Hermitian 3-diagonal matrix)
\[
J^* =
 \begin{pmatrix}
  B_{0}^* & J_1^* & 0 &    \\
  J_{1}^* & B_{1}^* & J_2^* & 0  \\
   0  &  J_2^* & B_2^* & J_3^*     \\
   &   &  \ddots &    \ddots  \\
    & & \dots & J_N^* & B_N^*
\end{pmatrix}
\]
with the coefficients
\be
B^*_n = B_{N-n}, \quad J_{n}^* = J_{N+1-n} . \lab{BJ_star} \ee
Consider the corresponding monic orthogonal polynomials $P_n^*(x)$ satisfying the recurrence relation
\be
P_{n+1}^*(x) + B_n^* P_n^*(x) + U_n^* P_{n-1}(x) = x P_n^*(x) . \lab{star_rec} \ee
In view of  \re{J*J}, the matrix $J^*$ has the same eigenvalues $x_s, s=0,1,\dots, N$ as the matrix $J$.

Hence the polynomials $P_n^*(x)$ are orthogonal on the same finite
set of spectral points \be \sum_{s=0}^N P_n^*(x_s) P_m^*(x_s)
w^*_s = h_n^* \delta_{nm} \lab{dual_ort} \ee but with another set
of weights $w_s^*$. There is a remarkable relation between the
weights $w_s$ and $w_s^*$ \cite{BS}, \cite{Bor}, \cite{VZ1}: \be
w_s w_s^* = \frac{h_N}{(P_{N+1}'(x_s))^2}. \lab{ww*} \ee

Now, if $J$ is mirror-symmetric, that is $J^*=J$, we must have
$w_s^*=w_s$ and in view of \re{ww*} both sets of weights are equal
and given by \re{w_s_kap}. A Jacobi matrix $J$ with mirror
symmetry  hence  defines a spin chain with PST. Conversely, if the
matrix $J$ leads to a system with PST, the weights $w_s$ of the
associated OPs must be given by formula \re{w_s_kap}. From
\re{ww*}, it then follows that $w_s^*=w_s$. This in turn implies
that $J^*=J$, in other words, that $J$ is mirror-symmetric. Note
that in the mathematical literature,  the matrices with the
property $J^*=J$ are called persymmetric \cite{BG},
\cite{Gladwell}.

In summary, \re{pqc_3} is equivalent to the condition that the
matrix $J$  is mirror-symmetric or persymmetric.

That \re{xxM} and the mirror-symmetry of the matrix $J$ are
necessary and sufficient for perfect state transfer was shown in
\cite{Kay}.

As a result, the exploration of systems with perfect state
transfer proceeded customarily through the search for Jacobi
matrices with mirror symmetry. We have just seen that this
property follows automatically from the prescription \re{w_s_kap}.
This observation considerably simplifies the problem of finding
spin chains with PST as their determination from the weight
formula \re{w_s_kap} avoids the necessity to check the mirror
symmetry of $J$.

At this point, to complete the practical recipe for the
construction of spin chains with PST,  we only need to provide an
efficient algorithm to determine the matrix $J$ from the spectrum.

One such possible algorithm (to reconstruct persymmetric matrices
from spectral data) was given in \cite{BG}.  We here propose
another one which seems more efficient especially in view of the
"spectral surgery" procedure to be presented in the next section.

First, we notice that the polynomial $\chi_N(x)$ can be reconstructed from \re{pqc_3} by the ordinary Lagrange interpolation procedure \cite{JJ}.

Indeed, $\chi_N(x)$ has degree $N$ and takes $N+1$ prescribed
values $\pm 1$ at $N+1$ prescribed distinct points $x_0, x_1,
\dots, x_N$.

It follows that \be \chi_N(x) =  \sum_{s=0}^N (-1)^{N+s} \mathcal
{L}_s, \lab{chi_Lag} \ee where $\mathcal{L}_s$ are the standard
Lagrange polynomials \be \mathcal{L}_s ={\prod_{i=0}^N}'
\frac{x-x_i}{x_s-x_i} \lab{Lag_s} \ee (as usual, the symbol $'$
means that $i \ne s$ in the product \re{Lag_s}).  The monic
polynomial $P_N(x)$ is then obtained through the division of
$\chi_N(x)$ by the coefficient of its leading monomial.

We thus know explicitly two monic polynomials: $P_N(x)$ and
$P_{N+1}(x) = (x-x_0)(x-x_1) \dots (x-x_N)$.  Starting from these
polynomials it is possible to reconstruct step-by-step all the
orthogonal monic polynomials $P_n(x), \; n=N-1,N-2, \dots, 1$ by
the well known Euclidean algorithm.

Let us divide the polynomial $P_{N+1}(x)$ by the polynomial
$P_N(x)$: \be P_{N+1}(x) = q_N(x) P_N(x) + R_{N-1}(x),
\lab{div_N+1} \ee where $q_n(x) =x-\beta_N$ and $R_{N-1}(x)$ is
the residue of this division. By construction, we have on the one
hand  $\deg(R_{N-1}(x))=N-1$ and hence \be R_{N-1}(x) = \gamma_N
Q_{N-1}(x), \lab{R_N-1_Q} \ee where $Q_{N-1}(x)=x^{N-1} +
O(x^{N-2})$ is a monic polynomial of degree $N-1$.

On the other hand, we have the recurrence relation \re{recP} from
which we conclude that  $\beta_N = b_N, \; \gamma_N=u_N$ and
$Q_{N-1}(x)=P_{N-1}(x)$. We thus get the next monic orthogonal
polynomial $P_{N-1}(x)$ as well as the recurrence coefficients
$b_N, u_N$.

The same steps can then be repeated with the polynomials $P_N(x),
P_{N-1}(x)$ yielding the recurrence coefficients $b_{N-1},
u_{N-1}$ and the polynomial $P_{N-2}(x)$. Iteration will provide
all recurrence coefficients $b_n,\: n=0,1,\dots,N$ and $u_n, \:
n=1, 2, \dots, N$ together with the corresponding orthogonal
polynomials $P_n(x), \: n=1,2,\dots,N$.

From a physical point of view, the case $B_n=0$ is of special
interest because it corresponds to zero external magnetic field.
Jacobi matrices $J$ with zero diagonal entries $B_n=0$ correspond
to symmetric orthogonal polynomials satisfying the property
\cite{Chi} \be P_n(-x) =(-1)^n P_n(x). \lab{PP_sym} \ee
Conversely, \re{PP_sym} is equivalent to the condition $B_n=0$ in
the Jacobi matrix $J$ \cite{Chi}.

What is more important is that \re{PP_sym} is tantamount to the
following spectral properties \cite{Chi}:

(i) the eigenvalues are anti-symmetric, i.e.
\be
x_n =-x_{N-n}, \quad n=0,1, \dots, N \lab{anti_x} \ee
and

(ii) the weights are symmetric: \be w_s = w_{N-s}>0. \lab{sym_w}
\ee We can now apply these observations to the PST problem. Assume
that the eigenvalues $x_s$ satisfy the properties \re{xxM} and
\re{anti_x}.  Then, the weights $w_s$ constructed from formula
\re{w_s_kap} obviously satisfy \re{sym_w}.

For perfect state transfer to be achieved in spin chains with zero
magnetic field it is thus necessary and sufficient that the
one-excitation energies satisfy \re{xxM} and \re{anti_x}. A
similar result was obtained in \cite{Kay} using a method
associated to the inverse eigenvalue problem. Note that it is
assumed in \cite{Kay} that the matrix $J$ is mirror-symmetric. We
already observed that this assumption is superfluous as the mirror
symmetry of $J$ follow from \re{w_s_kap}.

\section{Two explicit examples}
As illustrations of how the constructive procedure we have
described can be applied, we present in this section two examples.
The first example is well known, the second one seems to be new.

Before proceeding, note that given a spectral data set $x_s, \:
s=0,1,\dots,N$ it is straightforward to obtain another set
satisfying  \re{xxM} by an affine transformation \be \tilde x_s =
\alpha x_s + \beta, \lab{til_x} \ee where $\alpha, \beta$ are
arbitrary real parameters. The corresponding Jacobi matrix will
then have for its entries \be \tilde B_n = \alpha B_n + \beta,
\quad \tilde J_n = \alpha J_n. \lab{til_J} \ee We can use this
freedom to choose the most convenient form of the spectral data.

In particular, it is always possible to choose the parameters
$\alpha,\beta$ so that the eigenvalues $\tilde x_s$  are integers
with alternating parity (all $x_{2s}$ even and all $x_{2s+1}$
odd).  We will use this observation in the following.

Consider first the uniform grid \be x_s = s -N/2, \quad
s=0,1,\dots N .\lab{uni_x} \ee Using \re{w_s_kap} we easily
reconstruct the weights $w_s$ to obtain the binomial distribution.
The corresponding orthogonal polynomials $P_n(x;N)$ are the
symmetric Krawtchouk polynomials and the entries of the matrix $J$
are \be B_n=0, \quad J_n^2 = \frac{n(N+1-n)}{4}.\lab{Kr_J} \ee
This example is well known and was in fact obtained in
\cite{Albanese} as a first example of inhomogeneous spin chain
with the perfect state transfer property.

The next example is less trivial and seems to have escaped notice.
Take a "hyperbolic" analogue of the uniform spectrum, i.e. \be x_s
= A (q^{-s+N/2} - q^{-N/2+s}), \quad s=0,1,\dots,N, \lab{hyp_x}
\ee where $0<q<1$ and $A$ are real parameters. These parameters
should be chosen so as to  enforce the condition that all
differences $x_{s+1}-x_s$ are positive odd integers. To that end,
it is sufficient to demand that all $x_s$ be integers with
alternating parity, say $x_0$ is even, $x_1$ odd, $x_2$ even and
so on.

To do this, let us notice that the spectral points \re{hyp_x}
satisfy the recurrence relation \be x_{s+1}+x_{s-1} =
(q+q^{-1})x_s, \quad s=1,2, \dots, N-1.\lab{rec_hyp} \ee In order
to ensure that all $x_s$ are integers with alternating parity we
need to require \be q+q^{-1}= K, \lab{qqk} \ee where $K=4,6,\dots$
is an arbitrary positive even integer in the case of even $N$ and
$K=6,10,14, 18, \dots$  in the case of odd $N$ (we avoid the case
$K=2$ because it leads to the degenerate case $q=1$ corresponding
to the uniform grid \re{uni_x}). Condition \re{qqk} means that $q$
is a special case of quadratic irrationality. The difference in
the ranges of $K$ for $N$ even and $N$ odd is explained by the
observation that for $N$ even the minimal distance between the
eigenvalues $x_s$ is $A(q^{-1}-q)$ while for $N$ odd the minimal
distance is $2A(q^{-1/2} - q^{1/2})$. The admissible values for
$K$ then easily follow from \re{rec_hyp}.

With the help of \re{w_s_kap} the weights are straightforwardly
found  (up to a normalization factor): \be w_s = (-1)^s \:q^{sN}
(1+q^{2s-N}) \: \frac{(q^{-N};q)_s (-q^{-N};q)_s}{(q;q)_s
(-q;q)_s}, \lab{w_hyp} \ee where
$$
(x;q)_n=(1-x)(1-xq) \dots(1-xq^{n-1})
$$
stands for the q-shifted factorial \cite{KLS}.

The weights \re{w_hyp} correspond to a special case of the q-Racah
polynomials that are orthogonal on the grid \re{hyp_x} \cite{KLS}.
Hence the entries $B_n, J_n$ of the Jacobi matrix $J$ can be
recovered from the known recurrence coefficients of the q-Racah
polynomials \cite{KLS}.

The spectrum $x_s$ is anti-symmetric $x_{N-s} =-x_s$, hence the
diagonal terms are absent $B_n=0$.  For the non-diagonal terms we
have from \cite{KLS} the expression \be U_n =J_n^2 = A^2
\frac{(1-q^{2n})(q^{2(n-N-1)}-1)}{(1+q^{2n-N-2})(1+q^{2n-N})}.
\lab{U_n_hyp} \ee It is easily seen that the coefficients
\re{U_n_hyp} are positive and satisfy the mirror symmetry
condition $U_{n}=U_{N+1-n}$.

\section{Spectral surgery and generating new PST chain}
Given a one-excitation spectrum $x_s, \:
s=0,1,\dots,N$ satisfying condition \re{xxM},  we have seen how to
construct the spin chain Hamiltonian with PST. The Jacobi matrix
which has the couplings and magnetic field strengths as its
entries is recovered  from the prescribed weights $w_s$ given by
\re{w_s_kap}, or, in an equivalent form \be w_s =
\frac{(-1)^{N+s}}{{\prod_{i=0}^N}'(x_s-x_i)}, \lab{w_s_eq} \ee
where it is assumed that $x_0<x_1<\dots<x_N$. By construction, all
weights are positive $w_s>0, \: s=0,1,\dots, N$. The weights $w_s$
in \re{w_s_eq} are defined up to an arbitrary positive common
factor which has no effect on the entries $J_i, B_i$ of the Jacobi
matrix $J$.

Consider the modified set of spectral data $x_1, x_2, \dots, x_N$
obtained by removing the first entry $x_0$. The corresponding
weights
$$
\tilde w_s = \frac{(-1)^{N+s}}{{\prod_{i=1}^{N}}'(x_s-x_i)}, \quad
s=1,2,\dots,N
$$
can be obtained from the initial weights in the following simple
manner: \be \tilde w_s = const (x_s-x_0) w_s . \lab{tw0} \ee This
procedure removes the eigenvalue $x_0$ and  preserves the
positivity of the resulting weights $w_s$.

Similarly, one can remove any fixed spectral point $x_j$: \be
\tilde w_s = const (x_s-x_j) w_s, \quad s=0,1,\dots, j-1, j+1,
\dots , N .\lab{wsj} \ee In this case, however, the new weights
will be positive only if either $j=0$ or $j=N$. In all other cases
the weights $w_s$ cannot be positive for all $s$.

Nevertheless, removing a pair of neighboring points $x_j, x_{j+1}$
through \be \tilde w_s = const (x_s-x_{j})(x_s-x_{j+1}) w_s.
\lab{wsjj} \ee  maintains positivity of the weights $\t w_s$ for
all $s$, preserves property \re{xxM} and thus provides a new
admissible Jacobi matrix $\tilde J$ which also generates perfect
state transfer.

This procedure of removing pairs of neighboring levels can
obviously be iterated.

The orthogonal polynomials $\t P_n(x)$ corresponding to the
weights \re{wsj} are obtained from the polynomials $P_n(x)$  by
the Christoffel transform \cite{Sz} \be \t P_n(x) =
\frac{P_{n+1}(x) - A_n P_n(x)}{x-x_j}, \lab{CT} \ee where
$$
A_n = \frac{P_{n+1}(x_j)}{P_n(x_j)}.
$$
The entries of the matrix $\tilde J$ are related to the entries of
the matrix $J$ by the well known formulas \cite{ZhS} \be \tilde
U_n = U_n \frac{A_n}{A_{n-1}}, \quad \tilde B_n = B_{n+1}
+A_{n+1}-A_n . \lab{CT_UB} \ee Formulas \re{CT} and \re{CT_UB} can
be applied iteratively in order to obtain new matrices $\t J$ with
perfect state transfer from a matrix $J$ with that property.

Assume now that the initial matrix $J$ describes a dynamics with
zero magnetic field, i.e. $B_n=0$.  The spectral points $x_s$
satisfy therefore the symmetry condition \re{anti_x}. In this case
one may remove a symmetric pair of boundary eigenvalues
corresponding to the first and last levels \be \t w_s = const \:
(x_s^2 - x_0^2) w_s \lab{sym_ww} \ee and find for the associated
polynomials $\tilde P_n(x)$ \cite{Sz} \be \tilde P_n(x) =
\frac{P_{n+2}(x) - K_n P_n(x)}{x^2-x_0^2}, \quad K_n =
\frac{P_{n+2}(x_0)}{P_n(x_0)}, \lab{VTP} \ee therefore obtaining a
new Jacobi matrix $\tilde J$ with zero magnetic field $\tilde
B_n=0$ and with \be \tilde U_n = U_n \frac{K_n}{K_{n-1}}.
\lab{VT_U} \ee If $N$ is odd, there is another possibility to
remove two neighboring levels, in this case from the middle of the
spectrum. Indeed, take \be \tilde w_s = const (x_s -
x_j)(x_s-x_{j+1})w_s = const (x_s^2 - x_j^2) w_s \lab{midd_VT_w},
\ee where $j=(N-1)/2$. Formulas \re{VTP} and \re{VT_U} remain
valid if one replace $x_0$ with $x_j$.

The idea of obtaining PST chains with $B_n=0$ from given ones was discussed in \cite{Wang}.
Our approach is much more explicit and exploits well known formulas from the theory of orthogonal polynomials.

As a simple (but already nontrivial) example, consider the removal
of  levels from the middle of the uniform spectrum  \re{uni_x}
with odd $N$. After a finite number $L$ of such transformations,
we obtain  spectral data of the form \ba &&x_s=-N/2, -N/2+1, \dots
-L-3/2,  -L-1/2,  \nonumber \\ && L+1/2, L+3/2, \dots, N/2-1, N/2.
\lab{gap_x} \ea For $L=0$ this spectral data set coincides with
\re{uni_x}. For $L=1,2,\dots <(N-1)/2$ we have two uniform grids
separated by a gap of length $2L+1$. This spectrum corresponds to
a spin chain with $B_n=0$ that has the perfect transfer of state
property  and that was introduced in \cite{Shi} and studied in
\cite{SJ}. We have shown in \cite{VZ2} that this model is
associated to the dual -1 Hahn polynomials.

Let us conclude by stressing here that  ALL spin chains with the
PST property  (not only those with $B_n=0$ as in \cite{Wang}) can
be obtained through such (iterated) spectral surgery from the PST
spin chain corresponding to the uniform grid \re{uni_x}.

\bigskip\bigskip
{\Large\bf Acknowledgments}
\bigskip

AZ thanks Centre de Recherches Math\'ematiques (Universit\'e de
Montr\'eal) for hospitality.  The authors would like to thank
Mathias Christandl and Maxim Derevyagin for stimulating
discussions. The research of LV is supported in part by a research
grant from the Natural Sciences and Engineering Research Council
(NSERC) of Canada.

\newpage

\bb{99}

\bi{Albanese} C.Albanese, M.Christandl, N.Datta, A.Ekert, {\it
Mirror inversion of quantum states in linear registers},  Phys.
Rev. Lett. {\bf 93} (2004), 230502.

\bi{BG} C. de Boor and G.~H.~Golub {\it The numerically stable
reconstruction of a Jacobi matrix from spectral data},  Lin. Alg.
Appl. {\bf 21} (1978), 245--260.

\bi{BS} C. de Boor, E.Saff, {\it Finite sequences of orthogonal
polynomials connected by a Jacobi matrix},  Lin. Alg. Appl. {\bf
75} (1986), 43--55

\bi{Bor} A.Borodin, {\it Duality of Orthogonal Polynomials on a Finite Set}, J. Stat. Phys.
{\bf 109}, (2002), 1109--1120.

\bi{Bose} S.~Bose, {\it Quantum communication through spin chain
dynamics: an introductory overview},  Contemp. Phys., {\bf 48},
(2007), 13 -- 30.

\bi{Chi} T. Chihara, {\it An Introduction to Orthogonal
Polynomials}, Gordon and Breach, NY, 1978.

\bi{Gladwell} G.M.L.Gladwell, {\it Inverse Problems in Vibration}, 2nd Edition, Kluwer, Dordrecht, 2005.




\bi{JJ} H.~Jeffreys and B.~S.~Jeffreys, {\it Methods of
Mathematical Physics},  3rd ed. Cambridge, England: Cambridge
University Press, 1988.

\bi{Kay} A.Kay, {\it A Review of Perfect State Transfer and its
Application as a Constructive Tool}, Int. J. Quantum Inf. {\bf 8}
(2010), 641--676;  arXiv:0903.4274.

\bi{Karbach} P.Karbach and J.Stolze, {\it Spin chains as perfect
quantum state mirrors},  Phys.Rev.A {\bf 72} (2005), 030301(R)

\bi{KLS} R. Koekoek,P. Lesky, R. Swarttouw, {\it Hypergeometric
Orthogonal Polynomials and Their Q-analogues},  Springer-Verlag,
2010.


\bi{Shi} T. Shi, Y.Li , A.Song, C.P.Sun,  {\it Quantum-state
transfer via the ferromagnetic chain in a spatially modulated
field}, Phys. Rev. {\bf A 71} (2005), 032309, 5 pages,
quant-ph/0408152.

\bi{SJ} N.Stoilova, J.Van der Jeugt, {\it An exactly solvable spin
chain related to Hahn polynomials},  SIGMA {\bf 7} (2011), 033, 13
pages.

\bibitem{Sz} G. Szeg\H{o}, Orthogonal Polynomials, fourth edition,  AMS, 1975.

\bi{Wang} Y.~Wang, F.~Shuang, H.~Rabitz,  {\it All possible
coupling schemes in XY spin chains for perfect state transfer},
Phys. Rev. {\bf A 84}, (2011) 012307, arXiv:1101.1156.

\bi{VZ1} L.Vinet and A.Zhedanov, {\it A characterization of
classical and semiclassical orthogonal polynomials from their dual
polynomials},  J. Comp. Appl. Math. {\bf 172} (2004), 41--48

\bi{VZ2} L.Vinet and A.Zhedanov, {\it Dual~$-1$~Hahn~polynomials
and~perfect~state~transfer}, arXiv:1110.6477.

\bibitem{ZhS} A.S. Zhedanov, {\it Rational spectral transformations
and orthogonal polynomials}, J. Comput. Appl. Math. {\bf 85}, no. 1
(1997), 67--86.

\eb

\end{document}